# NFS: A Hand Gesture Recognition Based Game Using MediaPipe and PyGame

Md. Rafid Islam
Software Engineering
Islamic University of Technology
Gazipur, Dhaka, Bangladesh
rafidislam@iut-dhaka.edu

Ratun Rahman
Software Engineering
Islamic University of Technology
Gazipur, Dhaka, Bangladesh
ratunrahman@iut-dhaka.edu

Akib Ahmed
Software Engineering
Islamic University of Technology
Gazipur, Dhaka, Bangladesh
akibahmed@iut-dhaka.edu

Rafsan Jany
Software Engineering
Islamic University of Technology
Gazipur, Dhaka, Bangladesh
rafsanjany@iut-dhaka.edu

*Abstract*— **This paper represents a game which interacts with humans via hand gesture movement. Nowadays, apps like this seem rare, and there seems to be a window opening for this kind of application to be more prevalent and useful in the near future. This application is based on hand gesture movement instead of being dependent on a keyboard and mouse. The main issue was to figure out how to utilize machine learning to make this application work as it should be. First, two games were selected one with a traditional controller and another with hand gesture method. Then these two games based on the difficulty to use, fun elements, gameplay, and replayability were compared. Though the difficulty increases but the other three aspects improve significantly. After going through all of that a conclusion can be drawn that people are more likely to play a simple hand gesture-based game.**

## I. INTRODUCTION

NFS a game based on hand gesture detection, and there are a couple of reasons why this game was chosen to work on. It shows how human-computer interaction may be utilized to create a proper game based on hand gesture recognition. This game is simple and easy to make. Less control and complexity are involved. It can be demonstrated, with ease, how a game can be created using HCI with the help of fundamental functions.

The main objective was to introduce hand gesture method to interact with the system. This method can help to interact with system without any kind of input device like keyboard or mouse.

The goal of this project would be to show how human-computer interaction concepts can be put into action to make a fully functional game which is easy to master and fun to play. The new concept of how to play a basic game like this with the help of human-computer interaction has helped evaluating a product significantly.

First, relevant works and examples of how other people have done this type of game or project were studied. After that, it is the analysis step. Relevant information was gathered after that. Next is the concept design phase some control options and models of cars or other objects were included. Then it is the implementation process and after that evaluation. Implementation phases consist of sensor-based tracking developing the prototype, creating graphics and environment setup.

## II. LITERATURE REVIEW

The game basically uses a hand posture estimation algorithm to track the movement. Based on the command gathered from the user's hand movement, the car moves and tries to avoid the obstacles. If the car hits any obstacle then the game is over. Basically, the command left and right is shown on the top right corner of the screen.

The primary concept behind this game is to construct a simple game after using a hand posture estimation algorithm. After that, simulate keystrokes on the keyboard to communicate between the two programs. Basic Knowledge of Python, OpenCV, and PyGame was needed to build the project.

There are several key fields involved with computer vision, including image processing, video capture and analysis, face identification, and object detection, but developing real-time applications requires a cross-platform library. This is where OpenCV, a C++-based program that has now been ported to Java and Python, comes in. It operates on a variety of operating systems, including Windows, macOS, Android, iOS, Linux. OpenCV is an open-source library for tasks including face identification, objection tracking, landmark detection, and more.

Although it is an excellent tool for computer vision, system development without consideration for the widest possible audience remains a major issue among entrepreneurs. There are also situations when both clients and developers are undecided about the level of success they wish to accomplish. As a result, both parties must be familiar with OpenCV when dealing with computer vision. [8]

Using the library, one will be able to do: Image reading and writing, Video capture and storage, Filtering and alteration of images are examples of image processing, Detecting features, Detecting objects in video or images, such as human body parts, automobiles, signs, and so on.

OpenCV can create pictures from scratch, draw images with code, record and store movies, process images, do feature detection, recognize particular objects and analyze films, and determine an object's orientation and motion. Robotics, medical, industrial automation, security, and transportation all use OpenCV. It may be used to determine a robot's position in robotics. Navigation, obstacle avoidance, and Human-Robot Interaction are all possible applications. It can aid patients with cell or tumor categorization and detection, 2D/3D segmentation, 3D organ reconstruction, and vision-guided robotic procedures in medicine.

In terms of industrial automation, it may assist in identifying stock flaws, barcodes and packaging, item sorting, document analysis, and many other tasks. This may be utilized in surveillance and biometrics for security, and it can also be used to detect driver vigilance and produce autonomous cars for transportation.

Pygame is a cross-platform collection of Python modules for making video games. It is a collection of computer graphics and sound libraries for the Python programming language. Pete Shinners created Pygame to take the role of PySDL. Pygame is well suited to the development of client-side applications that may be packaged as a standalone executable. Pygame has a drawing mechanism that enables the user to build and draw on an infinite number of canvasses. [9]

## III. METHODOLOGY

There were four phased that is used to develop this project. The phased includes analysis, concept design, implementation and evaluation.

### A. Analysis Phase

In this game the car will be controlled by hand gestures. The hand gesture will be playing the role of keystrokes. The car will be controlled by multiple orientations of the hand. The orientation is tracked by the hand landmark. There are 21 landmarks that have been detected to track the correct position of the hand and make a meaningful gesture by hand movement.

### B. Concept Design Phase

In the game, the player will have a car. The car can be controlled by the keyboard stroke. The car will face some obstacles like another car. The player has to control the car to survive and make a score. But additional features of hand gesture controlling has been added. The car can be moved right and left by the keyboard. The movement can be controlled by the right and left gestures. The index finger of both fingers will be detected and the system will take the decision which is the gesture of moving right and which is the gesture of the moving left. The gesture will be captured by the camera attached to the device. This game is applicable to the webcam. The webcam will capture the gesture of the player's hands and send it to the system control. The angle between the figure will be calculated to taking the right decision for the system.

### C. Implementation Phase

In this game, three Landmarks has been used 8, 5, and 0, i.e., INDEX_FINGER_TIP, INDEX_FINGER_MCP, and WRIST, from 21 landmarks respectively. Python programming language was used in this project. Next, the angle between these landmarks were calculated and based on that angle the directions or the inputs were detected. The libraries that were used are mediapipe, cv2, numpy, uuid, os, pynput.keyboard. The mediapipe will direct the hand form the webcam input and point the 21 landmark of the hand. "cv2" is being used for the real time right-left showing in the monitor. The matrix calculation is being done by the numpy library. "uuid" library is used for generating the unique ids. "Os" will help to communicate with the operating system. The keyboard strokes will be controlled through pynput.keyboard.

### D. Evaluation Phase

The left-right command is directed by the hand gesture. In the output screen, the left-right command will be shown. This command is equivalent to the keyboard And then the car can move left and right.

## IV. IMPLEMENTATION

Some step by step implementation techniques were used to fully develop the system. As mentioned earlier, NFS needs sensor based tracking, graphics creation and environment setup for implementation.

### A. Sensor Based Tracking

As mentioned earlier, 21 landmarks can be detected.

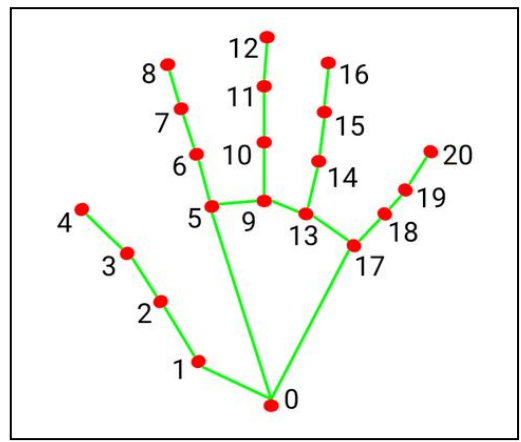

The landmarks are

0. Wrist
1. Thumb_CMC
2. Thumb_MCP
3. Thumb_IP
4. Thumb_TIP
5. Index_Finger_MCP
6. Index_Finger_PIP
7. Index_Finger_DIP
8. Index_Finger_TIP
9. Middel_Finger_MCP
10. Middel_Finger_PIP
11. Middel_Finger_DIP
12. Middel_Finger_TIP
13. Ring_Finger_MCP
14. Ring_Finger_PIP
15. Ring_Finger_DIP
16. Ring_Finger_TIP
17. Pinky_MCP
18. Pinky_PIP
19. Pinky_DIP
20. Pinky_TIP

As mentioned earlier, 0, 5, and 8 landmarks has been used in NFS.

## B. Create Graphics

Objects and elements were represented in 2D graphics. The score is shown but there are no visual effects.

## C. Environment

Camera is needed for the project or game to work. Pycharm was used for this Python development.

## V. RESULT AND EVALUATION

This game has been carried out by two testing methods mainly, user acceptance testing and usability testing. These testing techniques have helped to evaluate the game status and conditions as well as future goals.

## A. User Acceptance Testing

User acceptance testing (UAT) is mainly a black box technique used to analyze something by comparing factors like the expected results and actual results. Going thoroughly to the table helps to improve software quality and overlook the issue regarding the environmental factors in the real world [7].

Here, UAT was mainly focused on the core and critical components of the system. Table 1 shows the result from the test where 3 events were tested: detect hand gestures, car movement, and game mechanics.

TABLE I. GESTURE TESTING RESULTS

| Event | Expected Results | Actual Result |
|---|---|---|
| Detect Hand Gesture | Hand gesture will be recognized and work smoothly | Hand gesture recognition is working fine |
| Car Movement | Swiping hand towards a direction will work as a command to move the vehicle | The vehicle movement is working perfectly and precisely |
| Game Mechanics | Game will start and will end when the vehicle crash with other vehicles | There is no start menu and the game restart automatically after crush |

Based on the table, it shows that the hand detection and car movement are working fine. However, the game mechanics need some adjustment and improvement.

## B. Usability Testing

Usability testing is usually conducted to evaluate and analyze the product based on the real-life experience of the users. Here four factors have been chosen for this approach: difficulty to use, fun elements, gameplay, and replayability [6].

Eight questions were asked to 57 people who helped by replaying these questionnaires. A similar traditional game and NFS have been compared and significant changes have been found between them. People were asked to rate the answer on a 1 to 5 scale where 1 is the easier or less and 5 is the opposite.

Difficulties to use are a major concern for NFS as many people are not used to this hand gesture method. And figure 1 also shows the result. The traditional game is easier for people to use as it does not require additional complexity. However, people have also voted higher on hand gesture technique, mostly because this game does not require complex movement and control.

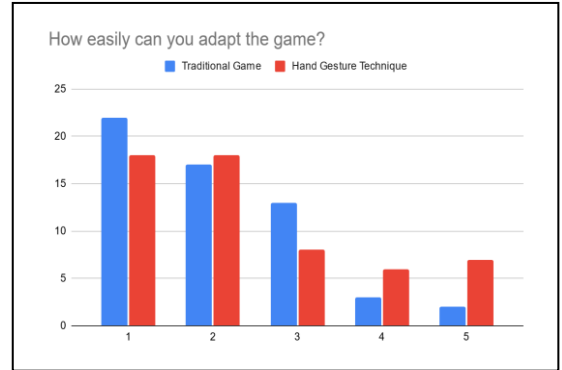

Figure 1. Difficulty to use the game

Fun elements, as the name suggests, focuses on the entertainment of the game. For any game, it is considered one of the most valuable elements. Figure 2 demonstrates that the game became more fun for hand gesture control than traditional control, mainly because of the uniqueness of the control.

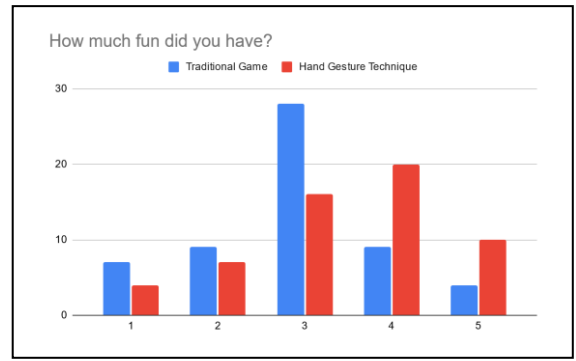

Figure 2. Entertainment of the game

Gameplay is the core component of the game, focusing on smoothness, controllability, reliability, and completeness. Based on figure 3, gameplay rating reduces for traditional methods while our hand gesture method has increased the rating.

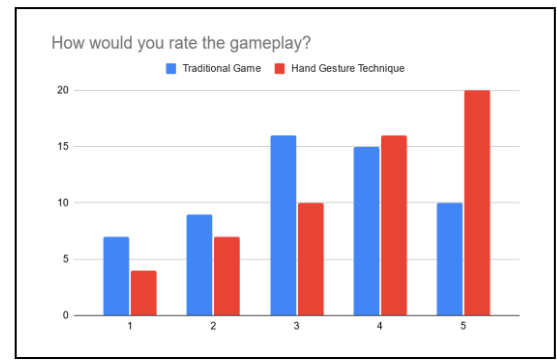

Figure 3. Gameplay

Replayability focuses on the tendency to play the game again. The more interesting the game is, the more people will tend to play it again. Figure 4 demonstrates that the hand gesture technique has more replayability than the traditional game method.

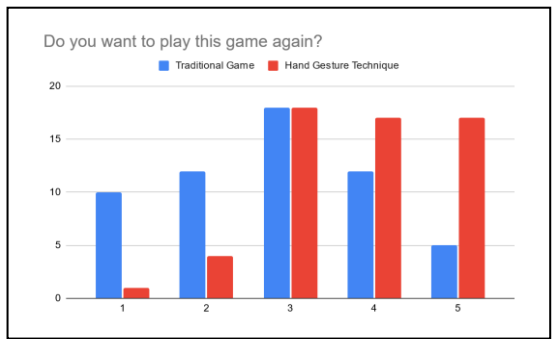

Figure 4. Replayability

Therefore, the difficulty increases but the other three factors improve significantly in hand gesture technique.

## VI. Conclusion

After the implementation and testing of the game, NFS has been developed based on the methodology proposed in this paper. The control using hand gestures has evaluated the game and has increased the game quality.

However, in this paper's development, there are some limitations that have been identified. The game itself lacks several visual elements including sounds and interfaces. However, this game was deliberately chosen as it is simple and easy to implement as well as has very few controls.

This concept can be applied to various other games as well including complex games and 3D games. However, for complex control, some features might not work as efficiently as NFS. But with better data training and gathering, the accuracy of the method can be increased.

Finally if given more time to do further development, several new concepts could have been added as well as new control features. This game could have been more interesting and challenging if there were a multiplayer element.

## References

[1] Khan, R.Z. and Ibraheem, N.A., 2012. Hand gesture recognition: a literature review. International journal of artificial Intelligence & Applications, 3(4), p.161.

[2] Fang, Y., Wang, K., Cheng, J. and Lu, H., 2007, July. A real-time hand gesture recognition method. In 2007 IEEE International Conference on Multimedia and Expo (pp. 995-998). IEEE.

[3] Rautaray, S.S. and Agrawal, A., 2011, December. Interaction with virtual game through hand gesture recognition. In 2011 International Conference on Multimedia, Signal Processing and Communication Technologies (pp. 244-247). IEEE.

[4] Zhang, X., Chen, X., Li, Y., Lantz, V., Wang, K. and Yang, J., 2011. A framework for hand gesture recognition based on accelerometer and EMG sensors. IEEE Transactions on Systems, Man, and Cybernetics-Part A: Systems and Humans, 41(6), pp.1064-1076.

[5] Garan, T.A. and Suaib, N.M., Hand Gesture Integration of 3D Virtual Rubik's Cube Using Leap Motion.

[6] Lewis, J.R., 2006. Usability testing. Handbook of human factors and ergonomics, 12, p.e30.

[7] Hambling, B. and Van Goethem, P., 2013, May. User acceptance testing: a step-by-step guide. Chippenham: BCS.

[8] An overview of opencv. Full Scale. (2022, April 12). Retrieved April 15, 2022, from https://fullscale.io/blog/opencv-overview/

[9] Pygame tutorial - javatpoint. www.javatpoint.com. (n.d.). Retrieved April 15, 2022, from https://www.javatpoint.com/pygame